\documentstyle[12pt,thmsa,sw20lart]{article}

\input{tcilatex}
\begin{document}

\title{Neutrino flavor mixing in an $SU\left( 3\right) $ symmetry for light
neutrinos}
\author{Riazuddin\thanks{%
email: ncp@comsats.net.pk} \\
Abdus Salam International Centre for Theoretical Physics Trieste, Italy\\
and\\
National Centre for Physics, Quaid-i-Azam University\\
Islamabad, Pakistan\thanks{%
Permanent address}}
\maketitle

\begin{abstract}
It is proposed that light neutrinos form a triplet in a global $SU(3)$
flavor symmetry. Within this framework, a symmetry principle predicts $\sin
^{2}\theta _{23}=0.5$ and gives the so called inverted neutrino mass
hierarchy, $\theta _{13}=0$ and a small deviation from $\pi /4$ for $\theta
_{12}$.
\end{abstract}

There is a compelling evidence\cite{r1} that neutrinos change flavor, have
non-zero masses and the neutrino mass eigenstates are different from weak
eigenstates. As such they undergo oscillations.

All neutrino data\cite{r1} with the exception of LSND anomaly\cite{r2} is
explained by three neutrino flavor oscillations with mass squared
differences and mixing angles having the following values\cite{r3}

\[
\Delta m_{solar}^{2}=\Delta m_{12}^{2}=\left( 8.1\pm 1.0\right) \times
10^{-5}\text{eV}^{2} 
\]
\[
\sin ^{2}\theta _{12}=0.30\pm 0.08 
\]
\begin{eqnarray*}
\Delta m_{atm}^{2} &=&\left| \Delta m_{12}^{2}\right| \simeq \left| \Delta
m_{23}^{2}\right| \\
&=&\left( 2.2\pm 1.1\right) \times 10^{-3}\text{eV}^{2}
\end{eqnarray*}
\[
\sin ^{2}\theta _{23}=0.50\pm 0.18 
\]
\[
\sin ^{2}\theta _{13}\leq 0.0047 
\]
The neutrino mixing angles are defined by the lepton mixing matrix \cite{r4}

\[
\left( 
\begin{array}{c}
\nu _{e} \\ 
\nu _{\mu } \\ 
\nu _{\tau }
\end{array}
\right) =U\left( 
\begin{array}{c}
\nu _{1} \\ 
\nu _{2} \\ 
\nu _{3}
\end{array}
\right) 
\]
The matrix $U$ is conveniently parametrized by three mixing angles $\theta
_{12},$ $\theta _{13},$ $\theta _{23}$ and three complex phases (two of
which are the so called Majorana CP odd phases) which we put equal to zero.

The purpose of this paper is to see whether one can find a symmetry
principle, which predicts $\sin^{2}\theta_{23}=0.5$. We show that it is
possible. Further one has the so called inverted neutrino mass hierarchy, $%
\theta_{13}=0$ and $\theta_{12}=\pi/4-\lambda$ where $\lambda$ is not fixed
by the proposed symmetry but is expected to be of order $\sin\theta_{c}%
\simeq0.22,$ $\theta_{c}$ being the Cabibbo angle.

Before we proceed further, let us remark that it is well known that a
neutrino mass matrix

\[
M_{\nu }=\left( 
\begin{array}{c}
0 \\ 
\cos \theta \\ 
\sin \theta
\end{array}
\begin{array}{c}
\cos \theta \\ 
0 \\ 
\sigma
\end{array}
\begin{array}{c}
\sin \theta \\ 
\sigma \\ 
0
\end{array}
\right) 
\]
with $\sigma <<1,$ which satisfies $\left( L_{e}-L_{\mu }-L_{\tau }\right) $
symmetry\cite{r5}, gives\cite{r6,r7}, after diagonalization the so called
inverted neutrino mass hierarchy, $\theta _{13}=0$ and $\theta _{12}=\pi /4.$
It also gives $\theta =-\theta _{23}$ but does not predict $\theta _{23}.$
The main handicap\cite{r8} is the prediction $\theta _{12}=\pi /4,$ which
does not seem to fit the data. The above mass matrix is naturally obtained
in the Zee model\cite{r9} or by a simple extension of the standard
electroweak gauge group to\cite{r7}

\[
G\equiv SU_{L}(2)\times U_{e}(1)\times U_{\mu }(1)\times U_{\tau }(1) 
\]

We propose that $\left( \nu _{e},\nu _{\mu },\nu _{\tau }\right) $ form a
triplet under a global $SU(3)$ neutrino flavor symmetry group and that
flavor changes are introduced through a Cabibbo like operator $[$reminiscent
of the Cabibbo formulation of weak interactions in quark flavor $SU(3)]$

\[
\cos \theta \left( O_{1}^{2}+O_{2}^{1}\right) +\sin \theta \left(
O_{3}^{2}+O_{2}^{3}\right) 
\]
Assuming further that it is $2\longleftrightarrow 3$ symmetric \cite{r10}
[again reminiscent of weak hadronic decays in quark flavor $SU(3)$] one
obtains

\[
\sin\theta=\cos\theta=\frac{1}{\sqrt{2}}, 
\]
giving $\theta=\pi/4.$ The resulting mass matrix, as noted in the previous
paragraph, will predict $\theta_{12}=\pi/4$ as well as $\theta_{23}=-\pi/4.$
Thus small perturbations have to be added to generate the required small
solar mass splitting and to avoid the prediction $\theta_{12}=\pi/4.$

Now two operators which have $2\longleftrightarrow 3$ symmetry are $%
O_{2}^{2}+O_{3}^{3}=-O_{1}^{1}$ and $O_{3}^{2}+O_{2}^{3}.$ Putting all the
terms together, we are led to the light neutrino mass matrix in the flavor
space 
\[
M_{\nu }=m_{0}\left( 
\begin{array}{c}
-\frac{2\varepsilon }{3} \\ 
\frac{1}{\sqrt{2}} \\ 
\frac{1}{\sqrt{2}}
\end{array}
\begin{array}{c}
\frac{1}{\sqrt{2}} \\ 
\frac{\varepsilon }{3} \\ 
\sigma 
\end{array}
\begin{array}{c}
\frac{1}{\sqrt{2}} \\ 
\sigma  \\ 
\frac{\varepsilon }{3}
\end{array}
\right) 
\]
Note that the tracelessness condition naturally emerges in our approch in
contrast to similar mass matrix studied in Refs. \cite
{hep-ph/0105212,hep-ph/0508053}. We expect the perturbation parameters

\[
\frac{\varepsilon }{3}\simeq \sigma \simeq \sin \theta _{c} 
\]
The diagonalization gives

\[
m_{2,1}\approx \left[ \pm 1+\frac{1}{2}\left( \sigma -\varepsilon /3\right)
\pm \frac{1}{8}\left( \varepsilon +\sigma \right) ^{2}\right] m_{0} 
\]
\[
m_{3}\approx -\left( \sigma -\varepsilon /3\right) m_{0} 
\]
Hence in the leading order 
\[
\Delta m_{12}^{2}\approx 2\left( \sigma -\varepsilon /3\right) m_{0}^{2} 
\]
\[
\left| \Delta m_{12}^{2}\right| \approx m_{0}^{2}\approx \left| \Delta
m_{32}^{2}\right| 
\]
\[
\theta _{13}=0 
\]
\[
\cos 2\theta _{12}=\frac{\frac{1}{2}\left( \sigma +\varepsilon \right) }{1+%
\frac{1}{8}\left( \varepsilon +\sigma \right) ^{2}}\simeq \frac{1}{2}\left(
\sigma +\varepsilon \right) 
\]
\[
\theta _{23}=-\theta =-\pi /4 
\]
In other words 
\[
\sin ^{2}\theta _{23}=0.5 
\]
and 
\begin{eqnarray*}
\sin ^{2}\theta _{12} &\approx &0.5-\frac{1}{4}\left( \sigma +\varepsilon
\right) \\
&\simeq &0.5-\lambda
\end{eqnarray*}
The data requires 
\[
\left( \sigma -\varepsilon /3\right) \simeq 0.038\pm 0.023 
\]
indicating a fine tuning between the values of $\sigma $ and $\varepsilon
/3, $ each of which is of order $\lambda .$ But fine tuning is a problem
common to Yukawa couplings even in charged sector, the solution of which has
so far eluded us.

To conclude we have found a symmetry $\left( 2\longleftrightarrow 3\right) $
within the framework of a global $SU(3)$ neutrino flavor symmetry, which
predicts $\sin ^{2}\theta _{23}=0.5$ and gives the so called inverted
neutrino mass hierarchy, $\theta _{13}=0$ and a deviation of $\theta _{12}$
from $\pi /4.$ The amount of this deviation is not predicted by the symmetry
but is likely to be of order $\lambda =\sin \theta _{c}\simeq 0.22,$ $\theta
_{c}$\cite{r11}.

{\bf Acknowledgments}

The author wishes to thank Dr. K. R. Sreenivasan for hospitality at the
Abdus Salam International Centre for Theoretical Physics and Dr. Luis
Alvariz-Gaume and the Theory Group for that at CERN, the two places where
this work was performed. He also acknowledges a research grant provided by
the Higher Education Commission of Pakistan to the author as a Distinguished
National Professor.

\end{document}